\documentclass[twocolumn,superscriptaddress,amsmath,ammsymb]{revtex4-1}
\usepackage{epsfig}
\usepackage{amsmath}
\usepackage{amssymb}
\usepackage{graphicx}
\usepackage{dcolumn}
\usepackage{bm}
\usepackage[usenames]{color}
\usepackage[breaklinks]{hyperref}
\usepackage{soul}
\usepackage{ulem}
\hypersetup{colorlinks=true,allcolors=blue}

\begin{document}
	
\title{Electron-transverse acoustic phonon couplings in three-dimensional pentatellurides} 
\author{Rui Min} 
\affiliation{School of Science, Jiangnan University, Wuxi 214122, China}
\author{Fuxiang Li} 
\affiliation{School of Physics and Electronics, Hunan University, Changsha 410082, China}
\author{Yi-Xiang Wang}
\email{wangyixiang@jiangnan.edu.cn}
\affiliation{School of Science, Jiangnan University, Wuxi 214122, China}

\date{\today}

\begin{abstract} 
Transverse acoustic (TA) phonon waves are analogous to electromagnetic waves and can carry a certain angular momentum.  
In this paper, we study the electron-TA phonon couplings in three-dimensional pentatellurides  and explore the conditions under which the TA phonon condensation is stable.  
We analyze the Lindhard response function, phonon softening, mean-field parameters, and renormalized dispersions, on the basis of which the phase diagrams of the electron-phonon couplings in ZrTe$_5$ and HfTe$_5$ are calculated.  
The phase diagrams show that, if the chemical potential lies near the Weyl nodes, the TA phonon condensation will dominate and lead to the shear strain wave phase.  
We further reveal that when the wave vector of the particular phonon mode is smaller, the critical coupling strength will be weaker for the phonon condensation, which thus favors the condensation phase. 
\end{abstract} 

\maketitle

\section{Introduction} 

In solid-state systems, the smooth lattice deformation or charge-neutralizing ionic background motion can give rise to collective excitations that are called phonons. 
When phonons are coupled to electrons, the so-called electron-phonon couplings represent a type of basic many-body interactions and form the basis of mechanisms for the conventional BCS superconductivity and Peierls transition~\cite{G.Gruner1988, G.Gruner2000}.  Even for the recently observed superconductivity in twisted bilayer graphene around the magic angle $\theta\simeq 1.05^\circ$~\cite{Y.Cao, P.Stepanov, Y.Saito}, the superconducting pairing can be understood from electron-phonon couplings~\cite{F.Wu, B.Lian}. 

Many previous electron-phonon coupling studies focused on the longitudinal acoustic (LA) phonon~\cite{B.Roy2014, Do, F.Buccheri}. 
The LA phonon can strongly couple with the electron-hole excitations, which  leads to the Peierls transition that is accompanied by the formation of the charge density wave (CDW), i.e., the periodic electron density distribution superimposed on the lattice periodicity.  
The CDW is rooted in the nesting property of the Fermi surface, where  nesting means that the two sections of the Fermi surface can be connected by the wave vector $Q$.  
Such a Peierls transition was successfully observed in one-dimensional (1D) organic and inorganic electronic conductors~\cite{D.Jerome} as well as (quasi-)1D semimetals~\cite{C.Yao}. 
Recently, the Peierls transition was suggested to be realized in three-dimensional (3D) topological semimetals under a magnetic field~\cite{M.Treschera, M.Trescherb, Z.Song} because the electrons will be quantized on 1D Landau bands and the corresponding Fermi surface have the nesting property.  

Different from the LA phonon wave, the transverse acoustic (TA) phonon wave is vectorlike and is similar to the electromagnetic wave.  
When the system has the rotational symmetry, the TA phonon is endowed with a definite chirality that is characterized by the polarization and can carry a certain angular momentum~\cite{L.Zhang}. 
Based on these facts, in Ref.~\cite{K.Luo}, the authors suggested that the TA phonon can also couple with electrons and lead to new kinds of the Peierls transition.  
They proposed topological semimetal systems in the quantum limit (QL) to study electron-TA phonon couplings.  Here QL is reached when all electrons occupy only the lowest zeroth Landau bands. 
More importantly, two new phases of the shear strain wave (SSW) and self-twisting wave (STW, also called the chiral CDW) emerge, which are caused by the linearly and circularly polarized TA phonon condensations, respectively~\cite{K.Luo}.  
However, under which conditions the TA phonon condensation is stable is still unclear, especially when it is competing with the LA phonon condensation.  
Here we try to address this question by studying the electron-TA phonon couplings in 3D pentatellurides.  
Three-dimensional pentatellurides, including ZrTe$_5$ and HfTe$_5$, represent a new type of topological material~\cite{H.Weng, B.Q.Lv, N.Wang}.  
At low temperatures, they feature very low carrier densities of about $10^{16}-10^{17}$ cm$^{-3}$~\cite{E.Martino, F.Tang, P.Wang, Galeski2020} 
and can reach the QL even under a weak magnetic field~\cite{W.Wu, Galeski2022, Z.Cai}, which thus  provides a good platform for exploring the physics of many-body interactions.  

In this paper, we use the effective $k\cdot p$ model to describe the low-energy excitations in 3D pentatellurides~\cite{W.Wu, Y.Jiang, J.Wang, Z.Cai, Y.X.Wang2023}.  Then we deal with the electron-phonon couplings using the mean-field strategy and solve the mean-field parameters using the gradient descent method.  
The Lindhard response function, phonon softening, mean-field parameters and renormalized dispersions are analyzed, on the basis of which we calculate the phase diagrams of the electron-phonon couplings in the parameter space  spanned by the magnetic field and the coupling strength.  
The results indicate that if the chemical potential lies around the Weyl nodes, TA phonon condensation will dominate and drive the system into the SSW phase; however, if the chemical potential lies near the band edge, LA phonon condensation will drive the system into the CDW phase.  
Moreover, we reveal that for the smaller wave vector of a particular phonon mode, the critical coupling strength of the phonon condensation will be weaker, and thus, the condensation phase is more stable. 
Our work will provide more insight into the electron-phonon couplings as well as the emerging new phases.

\section{Model}

The total model Hamiltonian describing electron-phonon couplings \textit{per unit volume} is written as 
$\hat H=\hat H_{\text{e}}+\hat H_{\text{ph}}+\hat H_{\text{e-ph}}$, where  the electron, phonon, and electron-phonon coupling terms are given as 
\begin{align}
&\hat H_{\text e}
=\frac{1}{V}\sum_{\alpha\beta,\boldsymbol k}{\cal H}_{\alpha\beta}(\boldsymbol k)
\hat c_{\alpha\boldsymbol k}^\dagger \hat c_{\beta\boldsymbol k}, 
\\
&\hat H_{\text{ph}}
=\frac{1}{V}\sum_{\lambda,\boldsymbol q}\hbar\omega_{\lambda}(\boldsymbol q)  
\hat b_{\lambda\boldsymbol q}^\dagger \hat b_{\lambda\boldsymbol q}, 
\\
&\hat H_{\text{e-ph}}
=\frac{1}{V}\sum_{\alpha\beta\lambda,\boldsymbol k\boldsymbol q}  
{\cal G}_{\alpha\beta\lambda}(\boldsymbol k,\boldsymbol q)  
\hat c_{\alpha\boldsymbol k+\boldsymbol q}^\dagger
\hat c_{\beta\boldsymbol k}(\hat b_{\lambda\boldsymbol q}
+\hat b_{\lambda,-\boldsymbol q}^\dagger), 
\end{align} 
respectively.  
Here $\hat c_{\alpha(\beta)\boldsymbol k}$ and $\hat b_{\lambda\boldsymbol q}$ are the electron and phonon annihilation operators, respectively, with $\alpha$ and $\beta$ being the band indices and $\lambda=x,y,z$ being the phonon polarization index.  
$\omega_{\lambda}(\boldsymbol q)=v_\lambda q$ is the acoustic phonon frequency, with $v_\lambda$ being the phonon velocity, and ${\cal G}_{\alpha\beta\lambda}(\boldsymbol k,\boldsymbol q)$ denotes the electron-phonon coupling matrix.  

In 3D pentatellurides, the ground states are believed to be located in proximity to the phase boundary between the strong topological insulator (TI) and weak TI~\cite{H.Weng}.  
Within the four-component basis $(|+\uparrow\rangle,|-\uparrow\rangle,|+\downarrow\rangle,|-\downarrow\rangle)^T$, the effective $k\cdot p$ model Hamiltonian describing the low-energy excitations is written as~\cite{Y.Jiang, Y.X.Wang2023, Z.Cai, J.Wang, W.Wu}
\begin{align}
{\cal H}_0(\boldsymbol k)=&\hbar v(k_x\sigma_z\otimes\tau_x+k_y I\otimes\tau_y)
+\hbar v_zk_z\sigma_x\otimes\tau_x
\nonumber\\
&+[M-\zeta(k_x^2+k_y^2)-\zeta_z k_z^2]I\otimes\tau_z, 
\end{align}
where $\sigma$ and $\tau$ are the Pauli matrices acting on the spin and orbit degrees of freedom, respectively.  $v$ and $v_z$ are the Fermi velocities, $\zeta$ and $\zeta_z$ are the coefficients of the quadratic terms, and $M$ is the Dirac mass.  Since the in-plane Fermi velocities are much larger than the out-of-plane ones, $v\gg v_z$, we take $v_z=0$.  

When a magnetic field $\boldsymbol B=B\boldsymbol e_z$ is applied to the 3D system, we adopt the symmetric gauge for the vector potential $\boldsymbol A=\frac{1}{2}(-yB,xB,0)$ and use the Peierls substitution to change the kinetic momentum to the canonical momentum  $\boldsymbol \pi=\boldsymbol p+e\boldsymbol A$.  
The orbital effect of the magnetic field will quantize the in-plane electron motion on 1D Landau bands. 
Within the framework of the symmetric gauge, the orbital angular momentum of the electron can be well expressed~\cite{D.Yoshika}.  
For Landau level (LL) wave functions $|n,m\rangle$, with $n$ denoting the LL index and $m$ the subindex, the electron takes the orbital angular momentum $l_z^{\text e}=(m-n)\hbar$. 
If the system has continuous rotational symmetry $\hat C_z$, the total angular momentum $j_z=l_z^e+s_z$ must be conserved, where $s_z$ is the spin inherited from the basis. 
In addition to the orbital effect, the magnetic field can also induce the spin Zeeman splittings, which are expressed as 
\begin{align} 
{\cal H}_Z=-\frac{1}{2}g_1\mu_BB\sigma_z-\frac{1}{2}g_2\mu_BB\sigma_z\tau_z. 
\end{align}
Here $\mu_B$ is the Bohr magneton, and $g_1$ and $g_2$ denote the Land\'e $g$ factors. 
In the QL, the low-energy physics is dominated by the second and third orbits, $|-\uparrow\rangle$ and $|+\downarrow\rangle$.  
The corresponding dispersions of the zeroth Landau bands are given as 
\begin{align}
\varepsilon_s(k_z)=s(\zeta_zk_z^2-M')+\frac{1}{2}g_2\mu_BB,
\label{varepsilon0} 
\end{align}
where $M'=M-\frac{\zeta}{l_B^2}+\frac{1}{2}g_1\mu_BB$ is the redefined Dirac mass,  $l_B=\sqrt{\frac{\hbar}{eB}}$ is the magnetic length, and $s=\pm$ characterize the conduction and valence bands when $\zeta_z>0$ and the valence and conduction bands when $\zeta_z<0$.  

For both LA and TA phonons, the canonical coordinates are written as  $\hat X_{\lambda q_z}=\xi_{\lambda q_z}(\hat b_{\lambda q_z}+\hat b_{\lambda,-q_z}^\dagger)$, with $\xi_{\lambda q_z}=\sqrt{\frac{\hbar}{2M_i \omega_\lambda}}$ being the zero-point displacement amplitude and $M_i$ the ion mass in each unit cell.  
In the linearly polarized basis, the canonical coordinate of the left- (right-) handed TA phonon can be expressed as $\hat X_{\pm,q_z}=\frac{1}{\sqrt 2}(\hat X_{xq_z}\pm\hat X_{y q_z})$, which carries the orbital angular momentum $l_{z,\pm}^{\text{ph}}=\pm\hbar$~\cite{L.Zhang}.  When the system includes both the electron-phonon couplings and spin-orbit couplings, the electron spin $s_z$, electron orbital angular momentum $l_z^{\text e}$ and phonon orbital angular momentum $l_{z,\lambda}^{\text{ph}}$ will all be coupled  together.   
Thus, we obtain the selection rules between the phonon-assisted transition processes for the wave vector, LL subindex, and total angular momentum~\cite{K.Luo}: 
\begin{align}
k_z=k_z'+q_z, \quad m=m', \quad s_z=s_z'+l_{z,\lambda}^{\text{ph}}, 
\label{rules}
\end{align}   
Since we focus on the QL, the LL index $n=0$ does not appear in the selection rules.  

To express the electron-phonon coupling matrix, we choose the four-component basis, $\hat\Psi_{mk_zQ}=(\hat c_{\alpha m,k_z+Q/2}, \hat c_{\beta m,k_z+Q/2}, 
\hat c_{\alpha m,k_z-Q/2}, \hat c_{\beta m,k_z-Q/2})^T$.  Here $Q$ is the nesting wave vector, and the spin $s_z^\alpha>s_z^\beta$.  
For TA and LA phonons, the coupling matrices are written as~\cite{G.Gruner2000,K.Luo}
\begin{align}
&{\cal G}_{\alpha\beta,\pm}(Q)=iQ\xi_{\pm,Q}g_\pm\sigma_\mp
\delta(s_z^\alpha-s_z^\beta-l_{z\pm}^{\text{ph}}), 
\label{Gpm}
\\
&{\cal G}_{\alpha\beta,z}(Q) =iQ\xi_{zQ}(g_0\sigma_0+g_z\sigma_z)\delta(s_z^\alpha-s_z^\beta), 
\label{Gz}
\end{align}
respectively.  Here $g_\lambda$ denotes the electron-phonon coupling strength and the Pauli matrices $\sigma_\lambda$ span the pseudospin space. 
As the $g_0$ and $g_z$ terms play equivalent roles in determining the mean-field parameters, we simply take $g_0=0$.

\section{Methods}

To solve the electron-phonon couplings, we use the mean-field strategy, in which the mean-field order parameter matrix is defined as~\cite{G.Gruner1988},  
\begin{align}
\Delta_Q=&\big[(1-e^{i\phi_{xy}}\Delta_{-,Q})\sigma_++
(1+e^{i\phi_{xy}}\Delta_{+,Q})\sigma_-
\nonumber\\
&+\Delta_{zQ}\sigma_z\big]\tau_++\text{H.c.}
\end{align}
Here $\phi_{xy}=\phi_y-\phi_x=\text{arg}\frac{\langle \hat X_{y,q_z}\rangle}{\langle \hat X_{x,q_z}\rangle}$ is the phase difference between TA modes in the $x$ and $y$ directions, and the mean-field parameters are defined as 
\begin{align}
&\Delta_{\pm,Q}=2iQ\xi_{\pm,Q}g_\pm\langle \hat b_{\pm,Q}\rangle,
\\
&\Delta_{zQ}=2iQ\xi_{zQ}g_z\langle \hat b_{zQ}\rangle, 
\end{align}
where $\hat b_{\pm,Q}=\frac{1}{\sqrt2}(\hat b_{xQ}\pm i\hat b_{yQ})$, and the expectation values $\langle\hat b_{\lambda Q}\rangle=\langle\hat b_{\lambda,-Q}^\dagger\rangle$ are assumed. 
With the help of $\Delta_Q$, the mean-field Hamiltonian for $\hat H_{\text e}+\hat H_{\text{e-ph}}$ is written as
\begin{align}
\hat{\bar H}_{mQ}(\boldsymbol k)=\big[{\cal H}_m(k_z)+\Delta_Q\big]
\hat\Psi_{mk_zQ}^\dagger \hat\Psi_{mk_zQ}. 
\end{align} 
 
In 3D pentatellurides, inversion symmetry $\hat{\cal P}\hat{\cal H}_m(k_z) \hat{\cal P}^{-1}=\hat{\cal H}_m(-k_z)$, with the operator $\hat{\cal P} =\tau_x\otimes\sigma_z$, exists. 
In the electron-phonon coupling terms, the inversion symmetry requires that $g_+=-g_-^*$ and $g_z=g_z^*$.  Further for the mean-field parameters, we have $\Delta_{+Q}=-\Delta_{-,-Q}^*$ and $\Delta_{zQ}=\Delta_{z,-Q}^*$.  
Thus, we can define $\Delta_T=|\Delta_{\pm Q}|$ and $\Delta_L=|\Delta_{zQ}|$.  
Moreover, the inversion symmetry guarantees that the left- and right-handed TA phonons are condensed with the same amplitudes, which correspond to the linearly polarized TA modes and the phase difference $\phi_{xy}=\pi$.  

\begin{figure}
	\includegraphics[width=9cm]{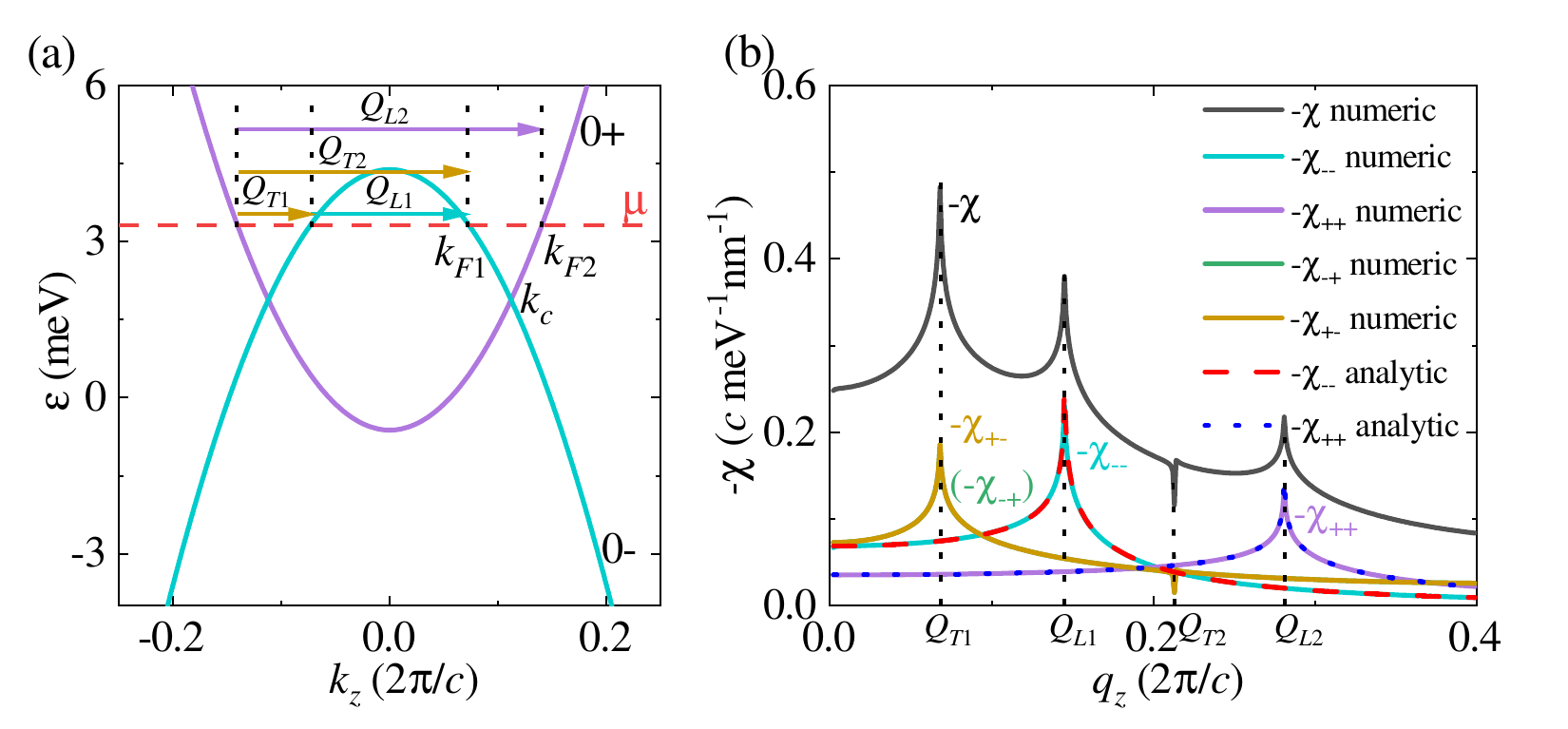}
	\caption{(Color online) (a) The two zeroth Landau bands with index $0\pm$ in 3D ZrTe$_5$.  The magnetic field is chosen to be $B=6.5$ T.  
	The horizontal dashed line denotes the position of the chemical potential $\mu$.  
	The arrows indicate the LA phonon with wave vectors $Q_{L1(2)}$, and the TA phonon with $Q_{T1(2)}$.  
	(b) The Lindhard response function $\chi$ and its components $\chi_{\alpha\beta}$ vs the wave vector $q_z$ at zero temperature.  
	For $\chi_{--}$ and $\chi_{++}$, the analytic results are also presented and show good consistency with the numerical ones.  
	Note that $\chi_{-+}$ is completely overlapped with $\chi_{+-}$. }  
	\label{Fig1}
\end{figure}

In the numerical calculations, we take the following steps to solve the mean-field parameter $|\Delta_{\lambda Q}|$ self-consistently: 
 
(i) Set an initial random value for $|\Delta_{\lambda Q}|$.

(ii) Diagonalize $\hat{\bar H}_{mQ}$ and calculate the ground-state energy \textit{per unit volume} $\bar E_g$ at zero temperature, 
\begin{align}
\bar E_g=&\frac{1}{V}\sum_{\boldsymbol k} 
[\varepsilon(k_z)-\mu]\Theta[\mu-\varepsilon(k_z)] 
\nonumber\\
&+\frac{1}{V}\sum_\lambda\frac{N_{p\lambda}M_iv_\lambda^2|\Delta_{\lambda Q}|^2}{g_\lambda^2},   
\end{align} 
where $\mu$ is the chemical potential and $\Theta(x)$ is the step function. 
The second term denotes the phonon energy expressed by $\Delta_{\lambda Q}$.   
The summations over $k_x$ and $k_y$ are calculated as $\sum_{k_x,k_y}=\frac{L_xL_y}{2\pi l_B^2}=N_L$, which gives the degeneracy factor for each LL and equals the phonon number $N_{p\lambda}=\hat b_\lambda^\dagger \hat b_\lambda$, meaning that a highly degenerate  electron gas requires the same number of phonons to be coupled.  We will take $L_z=c$ (the lattice constant in the $z$ direction).

(iii) Use the gradient descent method to find $|\Delta_{\lambda Q}|$, which can be achieved by minimizing $\bar E_g$.  In the $j$th iteration step, the order parameter is updated as $|\Delta_{\lambda Q}^{j}|=|\Delta_{\lambda Q}^{j-1}|-\gamma\frac{\delta{\bar E_g}(|\Delta_{\lambda Q}^{j-1}|)}{\delta|\Delta_{\lambda Q}^{j-1}|}$, with $\gamma$ being the step length. 

The convergence precision between two consecutive iterations is set to $10^{-8}$.  
If the convergence precision is reached, $\big||\Delta_{\lambda Q}^{j}|-|\Delta_{\lambda Q}^{j-1}|\big|<10^{-8}$, the iterations are completed; if it is not, repeat steps (ii)-(iii).  The results may converge to the ground state or some local minima.  To avoid the latter case, we need to choose several initial configurations for iterations, and select the state with the lowest $\bar E_g$ as its ground state.

\section{Strong topological insulator in ZrTe$_5$}

First, we study the strong TI state in 3D ZrTe$_5$.  
The strong TI features the band inversions in both the $x$-$y$ plane and $z$ direction.  
The model parameters are taken from the magnetoinfrared spectroscopy experiment~\cite{Y.Jiang}: $M=5$ meV, $v=6\times10^5$ m/s, $(\zeta,\zeta_z)=(100,200)$ meV nm$^2$, $g_1=-8$, and $g_2=10$, and the carrier density is fixed at $n_0=8\times10^{16}$ cm$^{-3}$.

\subsection{1D Weyl semimetal}

With the magnetic field set to $B=6.5$ T, the two zeroth Landau bands with index $0\pm$ are plotted in Fig.~\ref{Fig1}(a).   
We see that the two bands cross at $k_z=\pm k_c=\pm\sqrt{\frac{M'}{\zeta_z}}$, forming two valleys.  As the two crossing points have opposite chiralities, they can be regarded as a pair of Weyl nodes.  

The chemical potential $\mu$ is determined by the carrier density $n_0$ as
\begin{align}
n_0=\int_0^\infty d\varepsilon D(\varepsilon) f(\varepsilon-\mu)
+\int_{-\infty}^0 d\varepsilon D(\varepsilon) [f(\varepsilon-\mu)-1], 
\label{chempoten}
\end{align}
where $D(\varepsilon)$ is the density of states and $f(\varepsilon-\mu)=\frac{1}{\text{exp}[\beta(\varepsilon-\mu)]+1}$ is the Fermi distribution function with $\beta=\frac{1}{k_B T}$ being the inverse temperature. 
The charge neutrality is taken at the zero energy.  
By solving Eq.~(\ref{chempoten}), we obtain $\mu=3.33$ meV, as indicated by the dashed line in Fig.~\ref{Fig1}(a). 
Since $\mu$ intercepts the valence band and conduction band at $k_z=\pm k_{F1}$ and $k_z=\pm k_{F2}$, respectively, the Fermi surface includes four points and the system lies in the 1D Weyl semimetal (WSM) phase.

\subsection{Lindhard response function}

\begin{figure}
	\includegraphics[width=9.2cm]{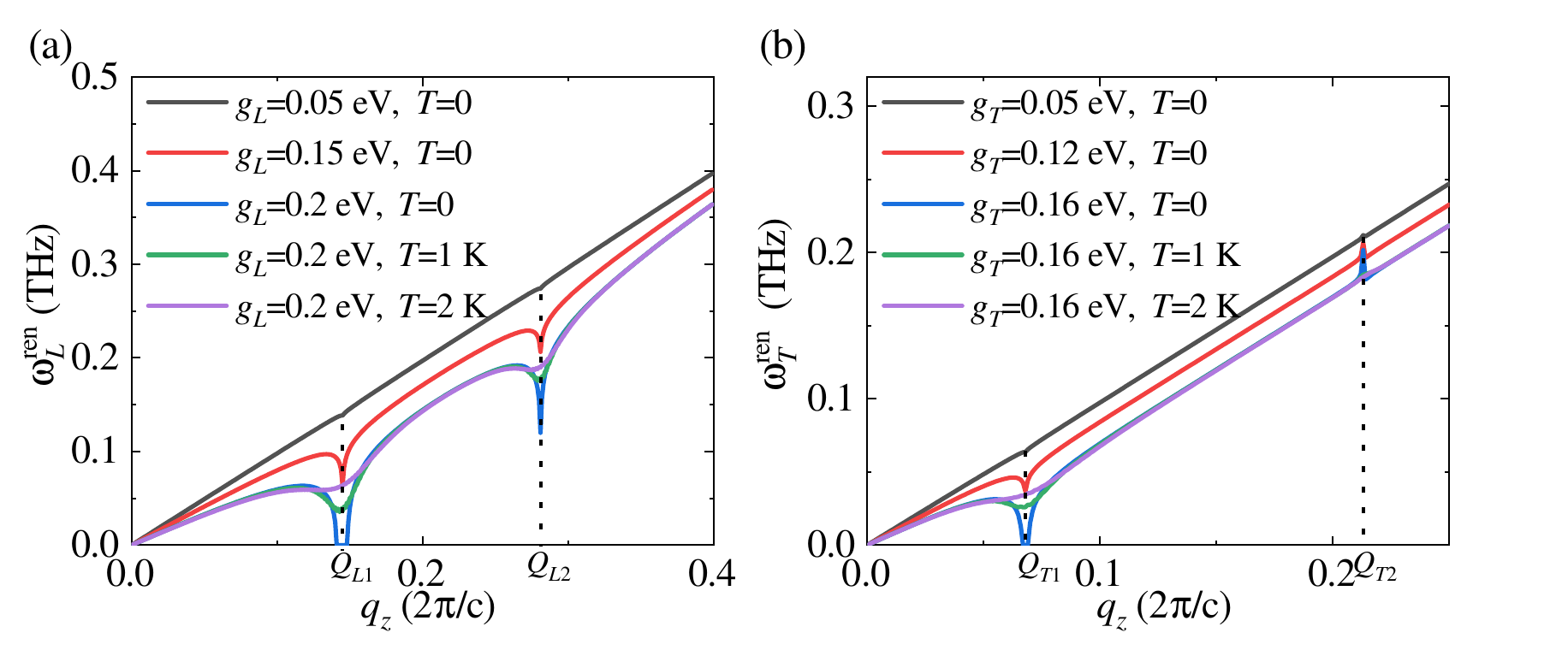}
	\caption{(Color online) The renormalized frequency $\omega_\lambda^{\text{ren}}$ for a set of the coupling strength and temperature $(g_\lambda,T)$, with (a) the LA phonon $\lambda=L$ and the TA phonon $\lambda=T$.  The wave vectors $Q_{L1(2)}$ and $Q_{T1(2)}$ are labeled by the dotted lines.  
	The model parameters are chosen to be the same as in Fig.~\ref{Fig1}, and the phonon wave speed is taken as $v_\lambda=10^3$ m/s.}  
	\label{Fig2}
\end{figure} 

To understand the role played by the phonon-assisted collective excitations in the system, we calculate the static Lindhard response function (also called the electron-hole polarizability), which is defined as~\cite{D.Jerome, B.Mihaila} 
\begin{align}
\chi(q_z)=\sum_{\alpha\beta}\chi_{\alpha\beta}(q_z), 
\end{align}
with its component  
\begin{align}
\chi_{\alpha\beta}(q_z)=\text{Re}\sum_{k_z}
\frac{f_{\alpha k_z}-f_{\beta,k_z+q_z}} {\varepsilon_\alpha(k_z)-\varepsilon_\beta(k_z+q_z)+i\eta}.  
\end{align}  
Here the band indices $\alpha,\beta=\pm$, and the Fermi distribution function $f_{\alpha k_z}=f[\varepsilon_\alpha(k_z)]$.  
The linewidth $\eta$ is introduced to avoid divergence and we take $\eta=0.1$ meV in the numerical calculations.  

In Fig.~\ref{Fig1}(b), we show the results of the Lindhard response function at zero temperature.  The components $\chi_{--}$ and $\chi_{++}$ are related to intervalley (intraband) transitions, in which the initial and final states have the same spin $s_z$.   
According to the selection rules in Eq.~(\ref{rules}), the LA phonons with vanishing orbital angular momentum can participate in such transitions.  
We see that the singularities of $\chi_{--(++)}$ are located at the wave vectors $Q_{L1(2)}=2k_{F1(2)}$.  
Here $Q_{L1}$ and $Q_{L2}$ are related to the valence and conduction bands, respectively.  
These singularities are rooted in perfectly nesting Fermi surfaces that are connected by $Q_{L1}$ and $Q_{L2}$ as 
$\varepsilon_-(k_z)=-\varepsilon_-(k_z+Q_{L1})+2\mu$ for $k_z$ around $-k_{F1}$ and 
$\varepsilon_+(k_z)=-\varepsilon_+(k_z+Q_{L2})+2\mu$ for $k_z$ around $-k_{F2}$, respectively.  
Moreover, when $\eta=0$, $\chi_{--}$ and $\chi_{++}$ can be calculated analytically, and the results are given as~(see Appendix A) 
\begin{align} 
\chi_{--(++)}=-\frac{c}{\zeta_zq_z}
\text{ln}\Big|\frac{q_z+Q_{L1(2)}}{q_z-Q_{L1(2)}}\Big|.  
\label{chiana}
\end{align}   
The analytic results of $\chi_{--}$ and $\chi_{++}$ are also plotted in Fig.~\ref{Fig1}(b), which show good consistency with the numerical ones. 

By contrast, the component $\chi_{+-}$ is related to intravalley (interband) transitions, in which the initial and final states have different spins $s_z$.    
Then the TA phonons with nonvanishing orbital angular momentum $l_{z\pm}^{\text{ph}}=\pm\hbar$ can participate in such transitions.  
We see that the singularities of $\chi_{+-}$ are located at $Q_{T1}=|k_{F1}-k_{F2}|$ and $Q_{T2}=k_{F1}+k_{F2}$.  
But $\chi_{+-}$ vanishes immediately when $q_z\neq Q_{T2}$, and the singularity shape at $q_z=Q_{T2}$ is quite different from the other three. 
This is because the Fermi surfaces connected by $Q_{T1}$ are still nearly nested as
$\varepsilon_+(k_z)\simeq-\varepsilon_-(k_z+Q_{T1})+2\mu$,  
for $k_z$ around $-k_{F2}$, but the Fermi surfaces connected by $Q_{T2}$ no longer have the nesting property, indicating that the $Q_{T2}$ phonon mode will not induce the Peierls instability. 

Within the random phase approximation of the electron-phonon couplings, the renormalized phonon frequency is determined by the Lindhard response function, which for the polarization $\lambda$ is given as
\begin{align} 
\omega_\lambda^{\text{ren},2}(q_z)
=\omega_\lambda^2+\sum_{\alpha\beta}\frac{2|{\cal G}_{\alpha\beta,\lambda}|^2
\omega_\lambda}{\hbar} \chi_{\alpha\beta}.     
\end{align}
After inserting the expressions for the electron-phonon coupling matrices ${\cal G}_{\alpha\beta,\lambda}$ in Eqs.~(\ref{Gpm}) and (\ref{Gz}), we have 
\begin{align}
\omega_\lambda^{\text{ren},2}(q_z) 
=\omega_\lambda^2+\frac{q_z^2 g_\lambda^2}{M_i} \sum_{\alpha\beta} \chi_{\alpha\beta}\delta(s_z^\alpha-s_z^\beta-l_{z,\lambda}^\text{ph}). 
\label{reomega}
\end{align} 

In Fig.~\ref{Fig2}, the renormalized phonon frequencies $\omega_\lambda^{\text{ren}}$ are plotted for a set of coupling strength and temperature $(g_\lambda,T)$.  
At zero temperature $T=0$, when $g_\lambda=0.1$ eV, both LA and TA phonon frequencies show minor variations.  
With increasing $g_\lambda$, $\omega_\lambda^{\text{ren}}$ will be reduced.  
The frequency reduction is most significant at $q_z=Q_{L1(2)}$ for a LA phonon and at $q_z=Q_{T1}$ for a TA phonon, which are dubbed phonon softenings~\cite{D.Jerome}. 
Note that for the TA phonon in Fig.~\ref{Fig2}(b), there is no phonon softening at $q_z=Q_{T2}$; just a spike is found.  
When the right-hand side of Eq.~(\ref{reomega}) is negative due to the strong $g_\lambda$, $\omega_\lambda^{\text{ren}}$ will be reduced to zero, which causes the Peierls instabilities. 
On the other hand, the phonon softenings are sensitive to a tiny temperature change; 
when $T=2$ K, the vanishing $\omega_\lambda^{\text{ren}}$ at $q_z=Q_{L1(T1)}$ will be smoothened to a finite value, meaning that the Peierls instabilities are broken. 

According to the above analysis, we suggest that in a 1D WSM, the Peierls transitions are caused by the $Q_{L1(2)}$ and $Q_{T1}$ phonon modes, but not by $Q_{T2},$ which will be further investigated in the following.

\subsection{Mean-field parameters}

\begin{figure}
	\includegraphics[width=9.2cm]{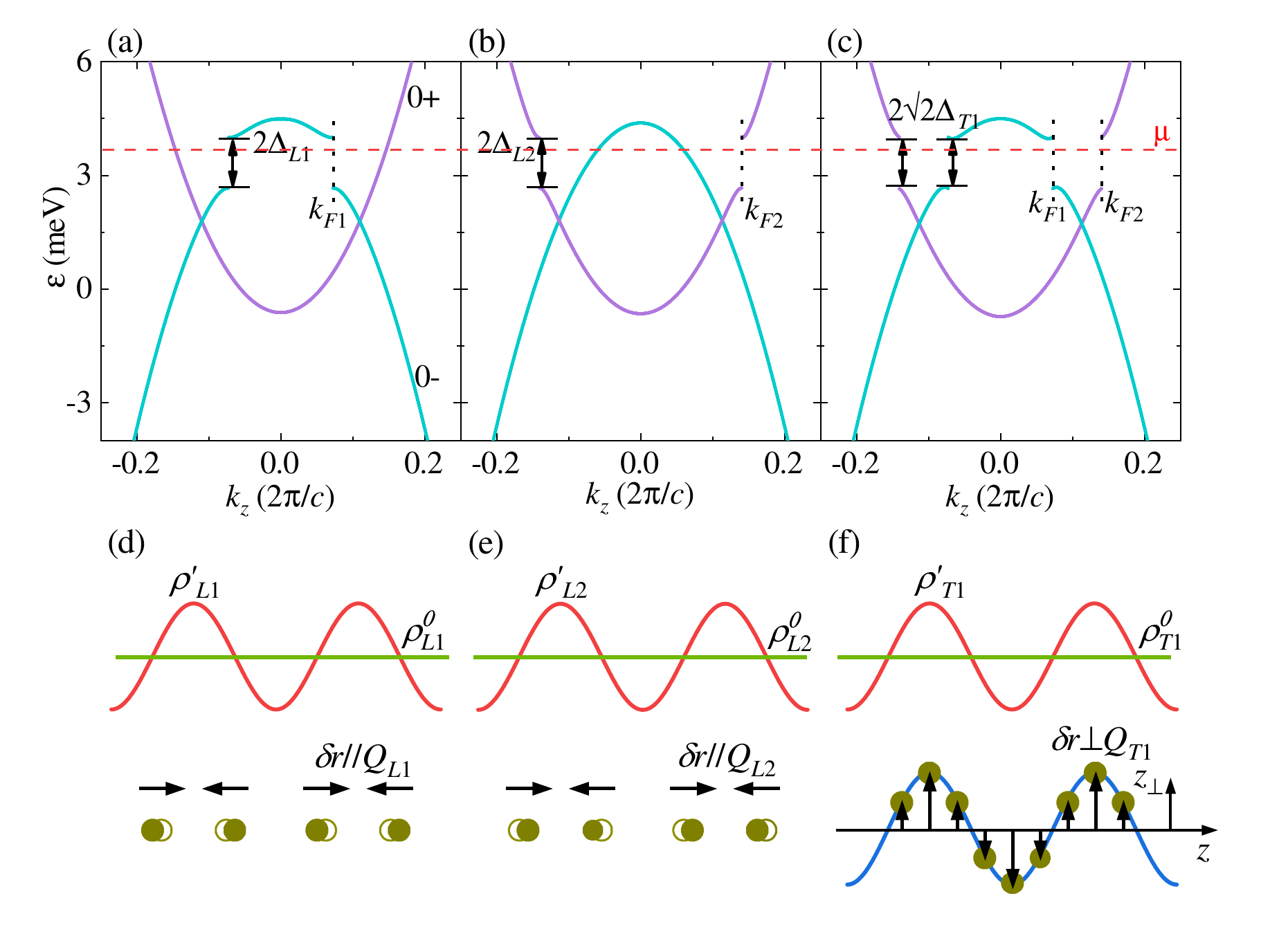}
	\caption{(Color online) (a)-(c) The renormalized zeroth Landau bands under the phonon condensation, with $\Delta_{L1}=0.67$ meV and $g_{L1}=0.246$ eV in (a), $\Delta_{L2}=0.67$ meV and $g_{L2}=0.366$ eV in (b), and $\Delta_{T1}=0.48$ meV and $g_{T1}=0.219$ eV in (c).
	(d)-(f) Schematics of the charge modulations and lattice distortions $\delta r$ in the CDW1, CDW2, and SSW phases.  Note that in (f), the unit vector on the $z_\perp$ axis is $\hat z_\perp=\frac{1}{\sqrt2}(\hat x-i\hat y)$.}  
	\label{Fig3}
\end{figure}

Next, we study the mean-field parameters $\Delta_{L1(2)}$ and $\Delta_{T1}$.  
Clearly, $\Delta_{L1(2)}$ is caused by $Q_{L1(2)}$ phonon condensation, and $\Delta_{T1}$ is caused by $Q_{T1}$ phonon condensation.  
We see that the mean-field parameters will renormalize the Landau bands.    
For $\Delta_{L1}$, only the valence band is renormalized.  When $k_z$ is around $\pm k_{F1}$, the renormalized dispersions for the valence band are given by~(see Appendix B)
\begin{align}
&\varepsilon_{\pm k_{F1}}=-M'+\frac{1}{2}g_2\mu_BB 
-\zeta_z\Big(k_z^2+\frac{Q_{L1}^2}{2}\mp k_zQ_{L1}\Big)
\nonumber\\
&\quad\mp\text{sgn}(k_z\mp k_{F1})
\sqrt{\zeta_z^2\Big(\frac{Q_{L1}^2}{2}\mp k_zQ_{L1}\Big)^2+\Delta_{L1}^2},
\label{epsilonkF1}
\end{align}
where sgn$(x)$ is the sign function. 
Similarly, for $\Delta_{L2}$, only the conduction band is renormalized. 
When $k_z$ is around $\pm k_{F2}$, the renormalized dispersions for the conduction band are given as 
\begin{align}
&\varepsilon_{\pm k_{F2}}=M'+\frac{1}{2}g_2\mu_BB 
+\zeta_z\Big(k_z^2+\frac{Q_{L2}^2}{2}\mp k_zQ_{L2}\Big)
\nonumber\\
&\quad\pm\text{sgn}(k_z\mp k_{F2}) 
\sqrt{\zeta_z^2\Big(\frac{Q_{L2}^2}{2}\mp k_zQ_{L2}\Big)^2+\Delta_{L2}^2}, 
\end{align}
In Figs.~\ref{Fig3}(a) and~\ref{Fig3}(b), the zeroth Landau bands are plotted.  
Around the chemical potential $\mu$, gaps of magnitude $2\Delta_{L1}$ and $2\Delta_{L2}$ are opened in the valence band and conduction band, respectively.  
Due to the gap opening, the charge will be redistributed along the wave vector in the $z$ direction.  
Then the system is driven into the CDW1 (CDW2) phase, with the electron density modulated as~\cite{F.Qin} 
\begin{align}
\rho_{L1(2)}(z)=\rho_{L1(2)}^0+\rho_{L1(2)}'\text{cos}[Q_{L1(2)}z+\phi_{L1(2)}].  
\end{align} 
Here $\rho_{L1(2)}^0$ denotes the constant electron density background, and $\rho_{L1}'=\frac{\Delta_{L1}}{\zeta_zQ_{L1}}
\text{arcsinh}[\frac{\zeta_zQ_{L1}(1-Q_{L1})}{2\Delta_{L1}}]$ and 
$\rho_{L2}'=\frac{\Delta_{L2}}{\zeta_zQ_{L2}}
\text{arcsinh}(\frac{\zeta_zQ_{L2}^2}{2\Delta_{L2}})$ are the oscillation amplitudes~(see Appendix B).  In Figs.~\ref{Fig3}(d) and~\ref{Fig3}(e), the schematics of the CDWs are plotted.  We see that the lattice distortions $\delta r$ in the CDW1 and CDW2 phases are parallel to wave vectors $Q_{L1}$ and $Q_{L2}$, respectively.  
For the large mean-field parameters, $\Delta_{L1(2)}\gg\zeta_zQ_{L1(2)}^2$, which are induced by strong couplings $g_{L1(2)}$, we have $\rho_{L1(2)}'\sim Q_{L1(2)}$. 

On the other hand, for $\Delta_{T1}$, both the conduction and valence bands will be  renormalized, as both bands are involved in intravalley transitions.  
When $k_z$ is around $\pm k_{F1}$, we have 
\begin{align}
&\varepsilon_{\pm k_{F1}}=\zeta_z\Big(\frac{Q_{T1}^2}{2}\pm k_zQ_{T1}\Big)
+\frac{1}{2}g_2\mu_BB\mp\text{sgn}(k_z\mp k_{F1})
\nonumber\\
&\quad\times\sqrt{\Big[\zeta_z\Big(k_z^2\pm k_zQ_{T1}+\frac{Q_{T1}^2}{2}\Big)-M'\Big]^2 +2\Delta_{T1}^2}, 
\end{align}
and when $k_z$ is around $\pm k_{F2}$, we have 
\begin{align}
&\varepsilon_{\pm k_{F2}}=-\zeta_z\Big(\frac{Q_{T1}^2}{2}\mp k_zQ_{T1}\Big)
+\frac{1}{2}g_2\mu_BB\pm\text{sgn}(k_z\mp k_{F2})
\nonumber\\
&\quad\times\sqrt{\Big[\zeta_z\Big(k_z^2\mp k_zQ_{T1}+\frac{Q_{T1}^2}{2}\Big)-M'\Big]^2
+2\Delta_{T1}^2}. 
\end{align} 
Figure~\ref{Fig3}(c) plots the zeroth Landau bands, where a gap of magnitude $2\sqrt2\Delta_{T1}$ is opened in the two bands.  
Different from the CDW formation under LA phonon condensation, such linearly polarized TA phonon condensation will generate a SSW with periodicity $\frac{2\pi}{Q_{T1}}$, and the system is called the SSW phase~\cite{K.Luo}.  In Fig.~\ref{Fig3}(f), the schematics show that the lattice distortions $\delta r$ in the SSW phase are perpendicular to the wave vector $Q_{T1}$ and form a wave in the shear strain plane $(z_\perp,z)$. 

\begin{figure}
	\includegraphics[width=9.2cm]{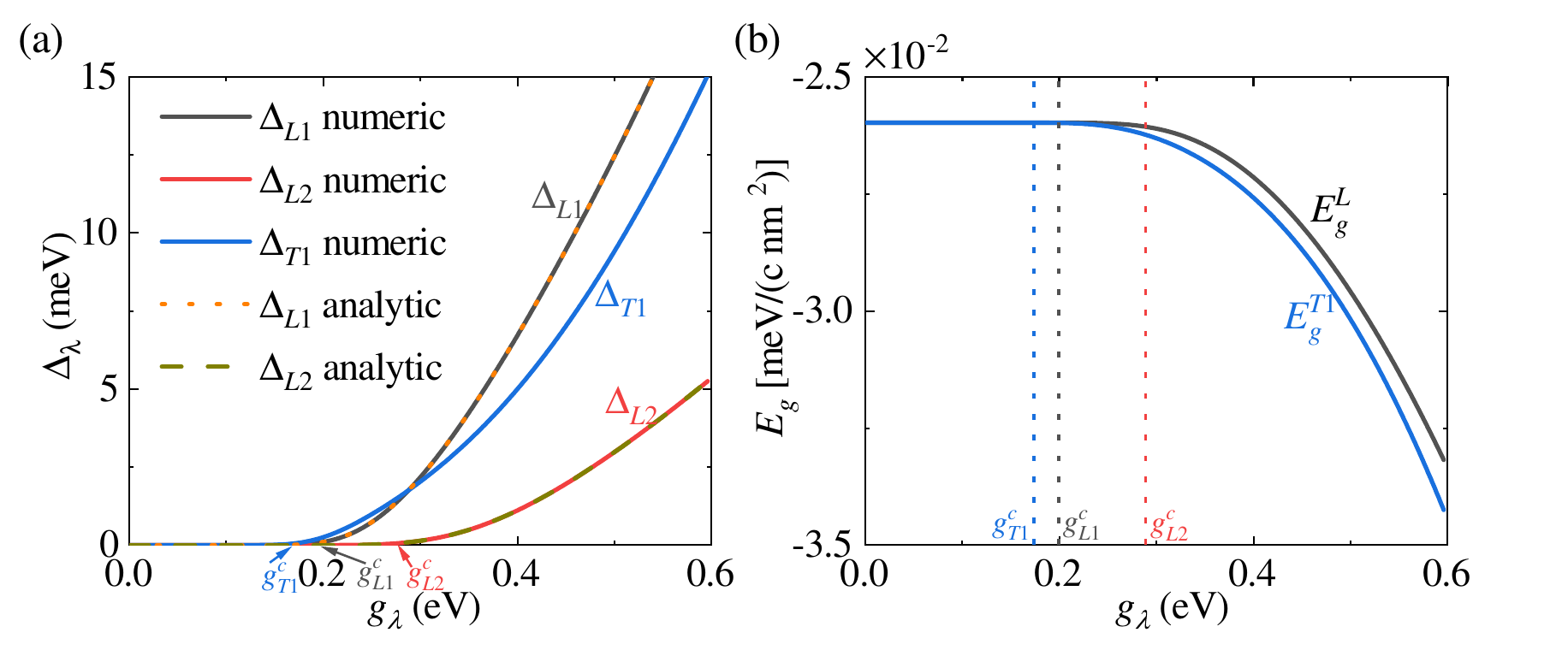}
	\caption{(Color online) (a) The mean-field parameters $\Delta_{L1}$, $\Delta_{L2}$, and $\Delta_{T1}$ vs the coupling strength $g_\lambda$.  
	For $\Delta_{L1}$ and $\Delta_{L2}$, the analytical results are also plotted and show good consistency with the numerical results. 
	The critical strengths are determined to be $g_{L1}^c=0.2$ eV, $g_{L2}^c=0.29$ eV, and $g_{T1}^c=0.18$ eV.  
	(b) The corresponding ground-state energies $\bar E_g^L$ and $\bar E_g^{T1}$. 
	The model parameters are chosen to be the same as in Fig.~\ref{Fig1}.}  
	\label{Fig4}
\end{figure}

We study the dependence of the mean-field parameters $\Delta_\lambda$ on the electron-phonon coupling strength $g_\lambda$.  
The numerical results are displayed in Fig.~\ref{Fig4}(a).  
We observe that in all mean-field parameters, the critical strengths $g_\lambda^c$ exists: When the coupling strength $g_\lambda<g_\lambda^c$, $\Delta_\lambda=0$, meaning that there is no phonon condensation and the system shows certain robustness to $g_\lambda$; 
when $g_\lambda$ is increased to $g_\lambda>g_\lambda^c$, $\Delta_\lambda$ becomes nonvanishing, and the spontaneous symmetry breaking in the system, which will drive the phonon condensation, exists. 
Note that the magnitudes of the critical $g_{L1}^c$ and $g_{T1}^c$ are consistent with those estimated from the Peierls transitions in Fig.~\ref{Fig2}. 

Actually, $\Delta_{L1}$ and $\Delta_{L2}$ can be solved analytically, with the results given as~(see Appendix B) 
\begin{align}
\Delta_{L1}=\frac{\zeta_zQ_{L1}(1-Q_{L1})}{2}\text{csch}
\Big(\frac{\zeta_zQ_{L1}M_iv^2}{c g_{L1}^2}\Big), 
\label{DeltaL1}
\end{align}
and
\begin{align}
\Delta_{L2}=\frac{\zeta_zQ_{L2}^2}{2}\text{csch}
\Big(\frac{\zeta_zQ_{L2}M_iv^2}{c g_{L2}^2}\Big). 
\label{DeltaL2}
\end{align} 
The analytical results for $\Delta_{L1}$ and $\Delta_{L2}$ are plotted in Fig.~\ref{Fig4}(a), which show good consistency with the numerical results.  
Equations.~(\ref{DeltaL1}) and~(\ref{DeltaL2}) indicate that 
(i) under weak $g_{L1(2)}<g_{L1(2)}^c$, the function csch$(x)\sim0$ and $\Delta_{L1(2)}$ vanishes;  (ii) a finite $\Delta_{L1(2)}$ requires that $g_{L1(2)}$ should be above the critical strength $g_{L1(2)}^c$;  (iii) when $g_{L1(2)}>g_{L1(2)}^c$, $\Delta_{L1(2)}$ increases with $g_{L1(2)}$.  
For $\Delta_{T1}$, although it cannot be obtained analytically, because it exhibits evolution similar to $\Delta_{L1}$ and $\Delta_{L2}$, we speculate that $\Delta_{T1}$ depends on $g_{T1}$ in a similar way.  
The existence of the critical $g_\lambda^c$ in electron-phonon couplings is analogous to several other physical factors in fermion systems, such as electron-electron interactions~\cite{B.Roy2017, Y.X.Wang2017, S.Rachel} and disorder~\cite{S.Bera,N.P.Armitage}. 

Now an important question arises: If all types of phonon condensation occur, which phase will dominate the ground state of the system, CDW1, CDW2, or SSW?  
Since in CDW1 and CDW2 the gap only opens in the conduction and valence bands, respectively, while in SSW the gap opens in both bands, we suggest that CDW1 and CDW2 may coexist to form the total CDW1+CDW2 phase but would compete with SSW to determine the ground state of the system.  
Note that in the SSW phase, the gaps in the conduction and valence bands have equal magnitudes; however, in the CDW1+CDW2 coexistence phase, the gaps are unequal.  
In Fig.~\ref{Fig4}(b), the ground-state energies are plotted for the cases of LA and TA phonon condensation. 
We find that when $g_\lambda>g_{T1}^c$, $\bar E_g^{T1}$ is lower than $\bar E_g^L$ even if both $\Delta_{L1}$ and $\Delta_{L2}$ are nonzero, meaning that the $\Delta_{T1}$ gap opening is more energetically favored than the $\Delta_{L1}$ and $\Delta_{L2}$ gap openings and therefore the system lies in the SSW phase.

\subsection{Phase diagram}

\begin{figure}
	\includegraphics[width=9.2cm]{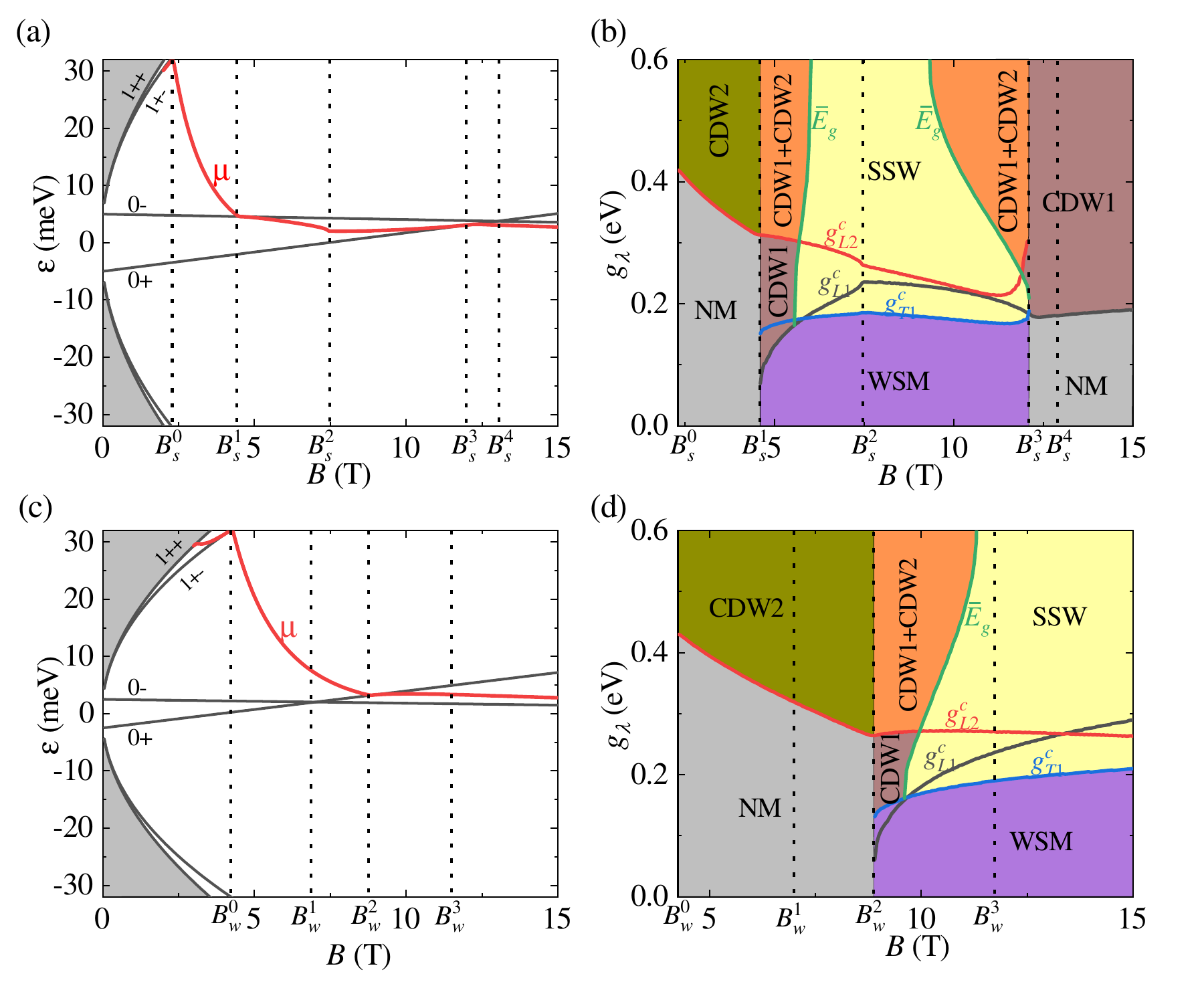}	
	\caption{(Color online) (a) and (c) The LL spectra and the chemical potential $\mu$ evolution with the magnetic field $B$ in ZrTe$_5$ and HfTe$_5$, respectively.  
	The characteristic fields $B_s^{0,\cdots,4}$ and $B_w^{0,\cdots,3}$ are denoted by the dotted lines. 
	In (a) the index $0+(-)$ denotes the conduction (valence) band; in (c), $0+(-)$ denotes the valence (conduction) band.  
	(b) and (d) The phase diagram of the electron-phonon couplings in the parametric space spanned by the magnetic field and the coupling strength $(B,g_\lambda)$ in ZrTe$_5$ and HfTe$_5$, respectively. 
	The different phases are shown by different colors.  
	The critical strengths $g_{L1}^c$, $g_{L2}^c$ and $g_{T1}^c$ are labeled by the black, red, and blue lines, respectively.  The phase boundaries judged by the ground-state energy $\bar E_g$ are denoted by green lines.} 
	\label{Fig5} 
\end{figure}

We further study the global phase diagram of the electron-phonon couplings, in which both LA and TA phonons are included.  
In determining the phase diagram, the position of the chemical potential $\mu$ under the condition of fixed carrier density plays an indispensable role.  
In Fig.~\ref{Fig5}(a), we show the evolution of $\mu$ with the magnetic field $B$ as well as the LL spectra.  
We see that $\mu$ exhibits a nonmonotonic variation.  More importantly, in the strong TI, the magnetic field will drive the two bands from crossing to separated, and several critical fields $B_s^{0,\cdots,4}$ exist~\cite{Z.Cai}: $B_s^0$ means that the system enters into the QL; $B_s^1=\frac{M-\mu}{e\zeta/\hbar-(g_1+g_2)\mu_B/2}$, $B_s^2=\frac{2\mu}{g_2\mu_B}$, and $B_s^3=\frac{M+\mu}{e\zeta/\hbar-(g_1-g_2)\mu_B/2}$ mean that $\mu$ meets the top of the valence band, the Weyl nodes, and the bottom of the conduction band, respectively; and $B_s^4=\frac{M}{e\zeta/\hbar-g_1\mu_B/2}$ means that the two bands are separated. 

In Fig.~\ref{Fig5}(b), the phase diagram is displayed in the parameter space spanned by the magnetic field $B$ and electron-phonon coupling strength $g_\lambda$.  
The phase boundaries are determined by the nonvanishing mean-field parameters $\Delta_\lambda$ together with the lowest ground-state energy requirement. 
We see that the phase diagram includes the phases with no phonon condensation, the normal metal (NM) and WSM, 
and the phases with phonon condensation, CDW1, CDW2, CDW1+CDW2, and SSW. 
Moreover, the phase diagram is separated by the critical fields into the following three regions. 

(i) When $B_s^0<B<B_s^1$, $\mu$ intercepts only the conduction band, and the system lies in the NM phase.  
The phonon wave vectors are given as $Q_{L1}=0$, $Q_{L2}\neq0$ and $Q_{T1}=0$.  
For the coupling strength above the critical value, $g_{L2}>g_{L2}^c$, the $Q_{L2}$ phonon mode will drive the system from the NM into the CDW2 phase.  
In this region, due to the steady decreasing of $\mu$, the critical $g_{L2}^c$ decreases with $B$.  

(ii) When $B_s^1<B<B_s^3$, $\mu$ intercepts the two zeroth Landau bands, and the system lies in the 1D WSM phase.  
Crossing the critical field $B_s^2$, the wave vectors turn from $0<Q_{L1}<Q_{L2}$ to $0<Q_{L2}<Q_{L1}$, both of which are accompanied by nonvanishing $Q_{T1}$.  
As a result, the $Q_{L1(2)}$ and $Q_{T1}$ phonon modes will compete to determine the ground state. 

Clearly, when $B$ is close to $B_s^2$, with $\mu$ located around the Weyl nodes, the SSW phase emerges as the energetically competitive ground state and spans a large area in the phase diagram.   
On the other hand, when $B$ is close to $B_s^{1(3)}$, with $\mu$ located near the top of the valence band (the bottom of the conduction band), the electron-phonon couplings will drive the system into the CDW1 (SSW) phase at first and then into the coexistence CDW1+CDW2 phase.  

(iii) When $B>B_s^3$, $\mu$ intercepts only the valence band, and the system also lies in the NM phase.  We have $Q_{L1}\neq0$, $Q_{L2}=0$, and $Q_{T1}=0$.  Similarly, for $g_{L1}>g_{L1}^c$, the $Q_{L1}$ phonon mode will drive the system into the CDW1 phase.  
As $\mu$ remains almost unchanged with $B$, the critical $g_{L1}^c$ line is quite flat in this region. 

The behavior in region (ii) can be more deeply understood from the perspective that the critical strength $g_\lambda^c$ is closely connected to the wave vector of the particular phonon mode.   
As seen in Eqs.~(\ref{DeltaL1}) and~(\ref{DeltaL2}), with a decreasing phonon wave vector $Q_{L1(2)}$, $\Delta_{L1(2)}$ will increase.  
For smaller $Q_{L1(2)}$, the weaker critical $g_{L1(2)}^c$ is required for the nonvanishing $\Delta_{L1(2)}$, leading to the $Q_{L1(2)}$ phonon condensation being more likely to occur.  
A similar relationship between $g_{T1}^c$ and $Q_{T1}$ is supposed to hold in $\Delta_{T1}$.  
Thus, if $\mu$ lies around the Weyl nodes, the wave vector of the $Q_{T1}$ phonon mode is the smallest, $Q_{T1}<Q_{L1},Q_{L2}$, and will drive the system into the SSW phase.  
By comparison, if $\mu$ lies near the top of the valence band (the bottom of the conduction band), the wave vector of the $Q_{L2}$ ($Q_{L1}$) phonon mode is the smallest, which then drives the system into the CDW phase.

\section{Weak topological insulator in HfTe$_5$}  

We study the electron-TA phonon couplings of the weak TI state in 3D HfTe$_5$.  
In the weak TI, the band inversions occur only in the $x$-$y$ plane and not in the $z$ direction.  
The model parameters are taken from the magnetoinfrared spectroscopy experiment~\cite{W.Wu}: $M=2.5$ meV, $(v,v_z)=(4.5,0)\times10^5$ m/s, $(\zeta,\zeta_z)=(120,-200)$ meV nm$^2$, $g_1=-6$, and $g_2=10$, and the carrier density is fixed at $n_0=1.4\times10^{17}$ cm$^{-3}$.  
As seen from the chemical potential evolution in Fig.~\ref{Fig5}(c), the magnetic field will drive the two bands from being separated to crossing, which is opposite of that in the strong TI.  
Now the critical fields $B_w^{0,\cdots,3}$ exist~\cite{Z.Cai, W.Wu}: $B_w^0$ means that the system enters the QL; $B_w^1=B_s^4$ means that the two bands cross, and $B_w^2=B_s^3$ and $B_w^3=B_s^2$ mean that $\mu$ meets the top of the valence band and the Weyl nodes, respectively.  

In Fig.~\ref{Fig5}(d), we display the phase diagram of the electron-phonon couplings in HfTe$_5$ and 
observe that the phase diagram is separated by the critical fields into two regions.  
(i) When $B_w^0<B<B_w^2$, the $Q_{L2}$ phonon mode will drive the system into the CDW2 phase. 
(ii) When $B>B_w^2$, the LA and TA phonons will compete with each other to determine the ground state.  
If $\mu$ lies near the Weyl nodes, the TA phonon-induced SSW phase spans a large region of parameter space; if $\mu$ lies near the top of the valence band, the LA phonon-induced coexistence CDW1+CDW2 phase will dominate. 
The features of these two regions are similar to those in Fig.~\ref{Fig5}(b).  
Moreover, the behavior in region (ii) can also be understood from the relationship between the critical coupling strength and the wave vector of a particular phonon mode.    
We note that region (iii) in Fig.~\ref{Fig5}(b) is absent here because in the weak TI, $\mu$ cannot move below the bottom of the conduction band even under a strong magnetic field~\cite{Z.Cai}.

\section{Discussion and Summary} 

We discuss the effect of the Dirac mass $M$ on the results.  
In ZrTe$_5$, the tiny Dirac mass behaves sensitively to external perturbations such as strain and  may even cause its ground state be classified as a weak TI~\cite{P.Zhang}.  
With decreasing $M$, the two crossing points of the zeroth Landau bands $k_z=\pm k_c$ will move to $k_z=0$, which suppress the wave vector $Q_{L1}$ as well as $Q_{T1}$ [see Fig.~\ref{Fig1}(a)].  In Fig.~\ref{Fig5}(b), the phase boundaries that are characterized by $B=B_s^1$ and $B=B_s^3$ will move towards the lower $B$, leading to the shrinking of the areas spanned by the CDW2 and SSW phases.  
Similar conclusions can also be found in Fig.~\ref{Fig5}(d) for the HfTe$_5$ case.  

In Figs.~\ref{Fig5}(b) and~\ref{Fig5}(d), the coexistence CDW1+CDW2 phase reminds us of the double CDWs that also refer to two charge modulations.  
The double CDWs were revealed in the multiband transition metal dichalcogenide NbSe$_2$ by using energy-dependent scanning tunneling microscopy measurements and resulted in spatial variations of the density amplitudes as well as energies~\cite{Y.W.Son, A.Pasztor}. 
However, in double CDWs, the two modulations are out of phase and occur at different energy levels, one at the Fermi energy and another below the Fermi energy, which, thus, differ from the conventional Fermi surface nesting picture.  

In Figs.~\ref{Fig5}(b) and~\ref{Fig5}(d), the phonon condensation requires that the electron-phonon coupling strength $g_\lambda$ lie within the range $0.2-0.6$ eV, which is comparable to the typical values of $0.1-1$ eV in real materials~\cite{F.Giustino}.  
Besides the strong coupling strength, the low temperature condition is also a prerequisite for phonon condensation (see Fig.~\ref{Fig2}). 
Nevertheless, a recent experiment reported that no signatures for CDW were captured in ZrTe$_5$ samples~\cite{Galeski2021}, which may be attributed to the weak coupling strength in the crystal or the high experimental temperature $T=2$ K. 

To summarize, in this work, we studied the electron-TA phonon couplings in 3D pentatellurides and revealed the stable conditions for TA phonon condensation and the resulting SSW phase.  Different from the LA phonon that couples Landau bands with the same spin, the TA phonon will couple Landau bands with different spins.  
Thus, the 1D WSM phase due to the Landau band crossings gives the necessary condition for TA phonon condensation.  
More importantly, if the chemical potential is modulated to lie around the Weyl nodes by changing the magnetic field, the wave vector of the particular TA phonon mode will be smaller. As a result, the TA phonon condensation is prone to occur, and the SSW phase emerges as the energetically competitive ground state. 
We emphasize that the appearance of the SSW phase is closely related to the linear polarized TA phonon to meet the inversion symmetry requirement in 3D pentatellurides.  
We expect that the above conditions for the SSW phase can also be extended to the STW phase, which may appear in a system with rotational symmetry~\cite{K.Luo}.  
In experiments, while the CDW phase can be detected by Raman scattering~\cite{Galeski2021} as well as x-ray diffraction~\cite{S.Gerber}, the detection of the SSW phase requires more study in the future.  
It is worth noting that in a recent experimental work on the Dirac semimetal EuAl$_4$~\cite{F.Z.Yang}, through x-ray scattering combined with second harmonic generation measurements, the transverse Peierls transition was identified as a possible origin of TA phonon softening.

\section{Acknowledgments}

This work was supported by the Natural Science Foundation of China (Grants No. 11804122, and No. 12275075), the National Key Research and Development Program of the Ministry of Science and Technology (Grant No. 2021YFA1200700), and the Fundamental Research Funds for the Central Universities from China.

\section{Data Availability} 

The data that support the findings of this article are openly available~\cite{data}.  The repository includes all relevant datasets and FORTRAN codes.

\section{Appendix}

\subsection{Calculation of $\chi_{--}$ and $\chi_{++}$}

For the Lindhard response function, the component $\chi_{--}$ that connects the two parts of the valence band is written as  
\begin{align}
\chi_{--}=c\int_{-\frac{1}{2}}^\frac{1}{2} dk_z 
\frac{f_{-,k_z}-f_{-,k_z+q_z}}{\varepsilon_-(k_z)-\varepsilon_-(k_z+q_z)},  
\tag{A1}
\label{chi--}
\end{align}
where the wave vector $k_z$ is in units of $\frac{2\pi}{c}$ and the Brillouin zone (BZ) is within the range $(-\frac{1}{2},\frac{1}{2})$.    
To complete the integral, we split it into two parts as 
\begin{align}
\chi_{--}=&c\int_{-\frac{1}{2}}^\frac{1}{2} dk_z
\Big[\frac{f_{-,k_z}}{\varepsilon_-(k_z)-\varepsilon_-(k_z+q_z)}
\nonumber\\
&-\frac{f_{-,k_z}}{\varepsilon_-(k_z-q_z)-\varepsilon_-(k_z)}\Big]. 
\tag{A2}
\label{A2}
\end{align}
For the second term in the brackets, the variable is redefined as $k_z+q_z\rightarrow k_z$, which is based on the fact that the integral is over all states in the BZ~\cite{B.Mihaila}. 
Inserting the expression for $\varepsilon_-$ in Eq.~(\ref{varepsilon0}) into Eq.~(\ref{A2}), we have
\begin{align}
&\chi_{--}=\frac{c}{\zeta_z q_z} 
\Big[\int_{-\frac{1}{2}}^{-k_{F1}} dk_z
\Big(\frac{1}{-q_z-2k_z}-\frac{1}{q_z-2k_z}\Big)
\nonumber\\
&\qquad\quad+\int_{k_{F1}}^\frac{1}{2} dk_z
\Big(\frac{1}{-q_z-2k_z}-\frac{1}{q_z-2k_z}\Big)\Big]
\nonumber\\ 
&=-\frac{c}{\zeta_z q_z}\text{ln}\Big|\frac{q_z+2k_{F1}}{q_z-2k_{F1}}\Big|. 
\tag{A3}
\end{align}  

Similarly, for the component $\chi_{++}$ that connects the two parts of the conduction band, we have 
\begin{align}
&\chi_{++}=c\int_{-\frac{1}{2}}^\frac{1}{2} dk_z  
\frac{f_{+,k_z}-f_{+,k_z+q_z}}{\varepsilon_+(k_z)-\varepsilon_+(k_z+q_z)} 
\nonumber\\
&=\frac{c}{\zeta_z q_z}\int_{-k_{F2}}^{k_{F2}}dk_z
\Big(\frac{1}{q_z+2k_z}-\frac{1}{-q_z+2k_z}\Big)
\nonumber\\ 
&=-\frac{c}{\zeta_z q_z}\text{ln}\Big|\frac{q_z+2k_{F2}}{q_z-2k_{F2}}\Big|. 
\tag{A4}
\end{align}

\subsection{Mean-field parameter $\Delta_{L1}$} 

Consider the LA phonon with wave vector $Q_{L1}=2k_{F1}$. 
In the basis $(\hat a_{1k_z},\hat a_{2k_z})^T=(\hat c_{\alpha m,k_z+Q/2},\hat c_{\alpha m,k_z-Q/2})^T$, the mean-field Hamiltonian $\hat H_{k_{F1}}$ around $k_{F1}$ is written as 
\begin{align*}
\hat H_{k_{F1}}=\begin{pmatrix}
\epsilon_{k_z}& -\Delta_{L1} e^{i\phi_{L1}}
\\
-\Delta_{L1} e^{-i\phi_{L1}}& -\epsilon_{k_z}
\end{pmatrix}-M''.
\tag{B1} 
\end{align*}
where $\epsilon_{k_z}=\zeta_z(\frac{Q_{L1}^2}{2}-k_zQ_{L1})$ and $M''=M'-\frac{1}{2}g_2\mu_BB-\zeta_z(k_z^2+\frac{Q_{L1}^2}{2}-k_zQ_{L1})$. 

To solve the $\hat H_{k_{F1}}$, we define a set of new bases 
\begin{align*}
&\hat\gamma_{1k_z}=U_{k_z} e^{-i\phi_{L1}/2}\hat a_{1k_z}-V_{k_z} e^{i\phi_{L1}/2}\hat a_{2k_z}, 
\\ 
&\hat\gamma_{2k_z}=V_{k_z} e^{-i\phi_{L1}/2}\hat a_{1k_z}+U_{k_z} e^{i\phi_{L1}/2}\hat a_{2k_z},   
\tag{B2} 
\end{align*}
with the constraint condition $U_{k_z}^2+V_{k_z}^2=1$. 
In the new basis, the Hamiltonian $\hat H_{k_{F1}}$ will be transformed into $\hat H_{k_{F1}}'$, which is given as  
\begin{align*}
&\hat H_{k_{F1}}'=\big[\epsilon_{k_z} (U_{k_z}^2-V_{k_z}^2)+2\Delta_{L1}U_{k_z}V_{k_z}\big]
(\hat\gamma_{1k_z}^\dagger\hat\gamma_{1k_z}
\\
&\quad-\hat\gamma_{2k_z}^\dagger\hat\gamma_{2k_z})
+\big[2\epsilon_{k_z} U_{k_z}V_{k_z}-\Delta_{L1}(U_{k_z}^2-V_{k_z}^2)\big] 
\\
&\quad\times(\hat\gamma_{1k_z}^\dagger\hat\gamma_{2k_z}+\hat\gamma_{2k_z}^\dagger\hat\gamma_{1k_z})-M''. 
\tag{B3}
\end{align*}  
When the nondiagonal terms in $\hat H_{k_{F1}}'$ become zero, $2\epsilon_{k_z} U_{k_z}V_{k_z}-\Delta_{L1}=0$, together with the above constraint condition, we have 
\begin{align*}
U_{k_z}^2=\frac{1}{2}\Big(1+\frac{\epsilon_{k_z}}{\sqrt{\epsilon_{k_z}^2+\Delta_{L1}^2}}\Big), 
\\
V_{k_z}^2=\frac{1}{2}\Big(1-\frac{\epsilon_{k_z}}{\sqrt{\epsilon_{k_z}^2+\Delta_{L1}^2}}\Big). 
\tag{B4}
\end{align*} 
After inserting $U_{k_z}$ and $V_{k_z}$ into the diagonal terms, we obtain the renormalized dispersions,  as shown in Eq.~(\ref{epsilonkF1}). 

The ground-state wave function is  
\begin{align*}
|\Phi_0\rangle
=\Big(\prod_{k_z}\hat\gamma_{1k_z}^\dagger\hat\gamma_{2k_z}^\dagger
\Theta[\mu-\varepsilon(k_z)]\Big)|0\rangle,  
\tag{B5}
\end{align*}
where $|0\rangle$ represents the vacuum.  The electron density is
\begin{align*}
\rho_{k_{F1}}(z)=\langle\Phi_0|\hat\Psi_{k_{F1}}^*(z)\hat\Psi_{k_{F1}}(z)|\Phi_0\rangle, 
\tag{B6}
\end{align*}
with  
\begin{align*}
\hat\Psi_{k_{F1}}(z)=\frac{1}{\sqrt{L_z}}\sum_{k_z}\big(\hat a_{1k_z}e^{ik_{F1}z}+\hat a_{2k_z}e^{-ik_{F1}z}\big).  
\tag{B7}
\end{align*}
By using the relations
\begin{align*}
&\langle\Phi_0|\hat\gamma_{1k_z}^\dagger\hat\gamma_{1k_z}|\Phi_0\rangle
=\langle\Phi_0|\hat\gamma_{2k_z}^\dagger\hat\gamma_{2k_z}|\Phi_0\rangle=1, 
\\
&\langle\Phi_0|\hat\gamma_{1k_z}^\dagger\hat\gamma_{2k_z}|\Phi_0\rangle
=\langle\Phi_0|\hat\gamma_{2k_z}^\dagger\hat\gamma_{1k_z}|\Phi_0\rangle=0, 
\tag{B8}
\end{align*}
after straightforward calculations, we have 
\begin{align*}
&\rho_{k_{F1}}
=\frac{1}{L_z}\sum_{k_z}\langle\Phi_0|
\big[U_{k_z}^2+V_{k_z}^2+2U_{k_z}V_{k_z}\text{cos}(Q_{L1}z+\phi_{L1})\big]
\\
&\qquad\quad\times
\hat\gamma_{2k_z}^\dagger\hat\gamma_{2k_z} 
|\Phi_0\rangle
\\
&=\rho_0+\int_{k_{F1}}^\frac{1}{2} dk_z 
\frac{\Delta_{L1}\text{cos}(Q_{L1}z+\phi_{L1})}
{\sqrt{\zeta_z^2\Big(\frac{Q_{L1}^2}{2}-k_zQ_{L1}\Big)^2+\Delta_{L1}^2}}
\\
&=\rho_0+\frac{\Delta_{L1}}{\zeta_zQ_{L1}}
\text{arcsinh}\Big[\frac{\zeta_zQ_{L1}(1-Q_{L1})}{2\Delta_{L1}}\Big]
\text{cos}(Q_{L1}z+\phi_{L1}), 
\tag{B9}
\end{align*}
The total electron density is obtained as 
\begin{align*}
\rho_{L1}(z)=\rho_{k_{F1}}+\rho_{k_{F2}}=\rho_{L1}^0+\rho_{L1}'\text{cos}(Q_{L1}z+\phi_{L1}).
\tag{B10}
\end{align*}
$\rho_{L1}^0$ denotes the constant electron density background, and $\rho_{L1}'$ gives the oscillation amplitude.  Since there is no gap opening around $k_{F2}$, the corresponding electron density contributes part of the constant background.  

The zero-temperature ground-state energy \textit{per unit volume} is 
\begin{widetext}
\begin{align}
&\bar E_g(\Delta_{L1})=\frac{N_L}{V} \Big[
\sum_{k_z\in(-\frac{1}{2},-k_{F1})}(\varepsilon_{-k_{F1}}-\mu)
+\sum_{k_z\in(k_{F1},\frac{1}{2})}(\varepsilon_{k_{F1}}-\mu) 
+\sum_{k_z\in(-k_{F2},k_{F2})}(\varepsilon_+-\mu)\Big]
+\frac{N_{pL1}M_iv^2\Delta_{L1}^2}{Vg_{L1}^2}. 
\tag{B11}
\end{align}
Note that in the brackets, the first term is equal to the second term due to inversion symmetry.  The derivation of $\bar E_g$ with $\Delta_{L1}$ is  
\begin{align}
\frac{\partial \bar E_g(\Delta_{L1})}{\partial\Delta_{L1}}
=&-\frac{2N_L}{V}\sum_{k_z\in(-\frac{1}{2},-k_{F1})} \frac{2\Delta_{L1}}{\sqrt{\zeta_z^2(Q_{L1}^2+2k_zQ_{L1})^2+4\Delta_{L1}^2}}
+\frac{2N_{pL1}M_iv^2\Delta_{L1}}{Vg_{L1}^2}
\nonumber\\
=&-\frac{2N_L}{V} \frac{\Delta_{L1}c}{\zeta_zQ_{L1}}
\text{arcsinh}\Big[\frac{\zeta_zQ_{L1}(1-Q_{L1})}{2\Delta_{L1}}\Big] 
+\frac{2N_{pL1}M_iv^2\Delta_{L1}}{Vg_{L1}^2},  
\tag{B12}
\end{align}
\end{widetext}
The condition of the lowest ground-state energy requires that $\frac{\partial \bar E_g(\Delta_{L1})}{\partial\Delta_{L1}}=0$, from which the analytical expression for $\Delta_{L1}$ is obtained, as shown in Eq.~(\ref{DeltaL1}).

\end{document}